\documentclass[11pt]{article}
\pdfoutput=1
\usepackage{graphicx,epsfig}
\usepackage{color}
\usepackage[colorlinks,citecolor=red,linkcolor=blue]{hyperref}
\usepackage{slashed}
\usepackage[left=2.5cm, right=2.5cm, top=2cm]{geometry}

\usepackage{epsfig,float,amsmath,amsfonts,amssymb,graphicx,bm,array,dcolumn}
\usepackage{bbold}

\newcommand{\Lagr}{\mathcal{L}}

\def\beq{\begin{equation}}
\def\eeq{\end{equation}}
\def\bea{\begin{eqnarray}}
\def\eea{\end{eqnarray}}
\def\bmat{\begin{pmatrix}}
\def\emat{\end{pmatrix}}
\newcommand{ \slashchar }[1]{\setbox0=\hbox{$#1$}   % set a box for #1
   \dimen0=\wd0                                     % and get its size
   \setbox1=\hbox{/} \dimen1=\wd1                   % get size of /
   \ifdim\dimen0>\dimen1                            % #1 is bigger
      \rlap{\hbox to \dimen0{\hfil/\hfil}}          % so center / in box
      #1                                            % and print #1
   \else                                            % / is bigger
      \rlap{\hbox to \dimen1{\hfil$#1$\hfil}}       % so center #1
      /                                             % and print / 
   \fi}                                             %
\def\etmiss{\slashchar{E}_{T}}
\def\to{\rightarrow}

\graphicspath{{./plots/}}

\begin{document}

\title{
  Gauge Singlet Vector-like Fermion Dark Matter, \\
   LHC Diphoton Rate and Direct Detection
}

\author{
%  \normalsize{
%
Shrihari~Gopalakrishna~\thanks{shri@imsc.res.in}~, 
Tuhin Subhra Mukherjee~\thanks{tuhin@imsc.res.in}~,  \\
\small{The Institute of Mathematical Sciences, HBNI,}  
\small{C.I.T Campus, Taramani, Chennai 600113, India.}
}

\maketitle

%%%%%%%%%%%%%%%%%%%%%%%%%%%%%%%%%%%%%%%%%%%%%%%%%%%%%%%%%%%%%%%%%%
\begin{abstract}

We study a gauge-singlet vector-like fermion hidden-sector dark matter model,  
in which the communication between the dark matter and the visible standard model sector is
via the Higgs-portal scalar-Higgs mixing,
and also via a hidden-sector scalar with loop-level couplings to two gluons and also to two hypercharge gauge bosons
induced by a vector-like quark. 
We find that the Higgs-portal possibility is stringently constrained to be small by the recent LHC di-Higgs search limits, 
and the loop-induced couplings are important to include. 
In the model parameter space, we present the dark matter relic-density, 
the dark-matter-nucleon direct-detection scattering cross-section,  
the LHC diphoton rate from gluon-gluon fusion,  
and the theoretical upper-bounds on the fermion-scalar couplings from perturbative unitarity. 
    
\end{abstract}

%\vfill\eject
%\tableofcontents
%\vfill\eject

%%%%%%%%%%%%%%%%%%%%%%%%%%%%%%%%%%%%%%%%%%%%%%%%%%%%%%%%%%%%%%%%%
\section{Introduction}
\label{Intro.SEC}

The nature of dark matter is yet to be established and many particle physics candidates in theories beyond the standard model (BSM) are being considered.
In this work, we add to the standard model (SM) a gauge-singlet ``hidden sector'' containing a vector-like fermion (VLF) dark matter $\psi$
and a scalar $\phi$.
We also add an $SU(3)_c$ (color) triplet, $SU(2)_L$ singlet, $U(1)_Y$ hypercharge $2/3$, vector-like quark (VLQ) $U$.
These states could be the lower energy remnants of a more complete theory which we do not need to specify here. 
The hidden sector is coupled to the standard model (SM)
via the ``Higgs-portal'' mechanism due to the $\phi$ mixing with the SM Higgs boson $h$,
and also via the loop-level couplings of the $\phi$ to two gluons
and the $\phi$ to two hypercharge gauge bosons induced by the $U$.

We analyze the large hadron collider (LHC) constraints on the model and find that the LHC di-Higgs channel imposes
a tight constraint on the scalar-Higgs mixing.
If the parameters are such that the Higgs-portal mixing is tiny, it becomes important to include the loop-induced couplings
of the $\phi$ to the SM induced by the VLQ that offers another mechanism of communication between the hidden sector and the SM
and generates the required size of the self-annihilation cross section that sets the dark matter relic density. 
Thus the presence of the VLQ is crucial for obtaining an acceptable phenomenology in the small Higgs-portal mixing limit. 
We present a scenario for which the scalar-Higgs mixing is tiny,
and only the loop induced couplings communicate between the hidden sector and the SM.
We show that the required values of the scalar-fermion Yukawa couplings are consistent with perturbative unitarity constraints
by considering the $\psi\bar\psi \to \psi\bar\psi$ process. 

The presence of the VLQ in the model affords a way to probe the model, and ongoing direct searches at the LHC are important. 
We briefly make contact with the extensive literature on searches for a VLQ at the LHC and the present constraints. 
Another signature of the model is a diphoton resonance signal due to $\phi$ production at the LHC via its digluon coupling
and subsequent decay into photons via its diphoton coupling, both of these induced by the VLQ at loop-level. 
We study the diphoton signature of the model in detail. 
In our model, we obtain expressions for the one-loop scalar-gluon-gluon ($\phi gg$) and scalar-photon-photon ($\phi\gamma\gamma$) effective couplings,
and explore the phenomenology including these couplings. 
We obtain expressions for some relevant decay modes of the $\phi$, 
present direct LHC constraints, the LHC gluon-gluon-fusion rate for $\phi$ production, 
and the LHC diphoton rate from $\phi$ decay for various total widths of the $\phi$.
For processes involving the dark matter, in addition to the Higgs boson contributions,
the new contributions here are the $s$-channel $\phi$ contribution to the $\psi\psi \to SM$ self-annihilation process
that sets the dark matter relic-density in the early universe,
and the $t$-channel $\phi$ contribution to the interaction of the dark matter with a nucleon that leads to a direct-detection signal. 
We compute the dark matter relic-density and the dark-matter-nucleon direct-detection cross section for this model.
Indirect detection of the dark matter via cosmic ray observables is another potential probe of the model,
which we do not purse in this study but leave for future work. 

We summarize next other studies in the literature that have some overlap with our work.
Ref.~\cite{Low:2011gn} studies loop induced couplings of a singlet scalar to electroweak gauge bosons. 
Precision electroweak observables, and scalar and Higgs phenomenology at the LHC with a singlet scalar and VLFs present are analyzed in
Ref.~\cite{Dolan:2016eki}. 
Singlet scalar decays to electroweak gauge bosons and to di-Higgs is studied in Ref.~\cite{Buttazzo:2015bka}. 
An analysis of a gauge-singlet fermionic dark matter in the Higgs portal scenario
with significant $\phi\leftrightarrow h$ mixing is carried out for example in 
Refs.~\cite{LopezHonorez:2012kv,Baek:2012se,Fedderke:2014wda,Gopalakrishna:2009yz,Kim:2006af,Kim:2008pp}.
The phenomenology of a singlet scalar coupled to
VLFs in the context of the earlier 750 GeV diphoton excess~\cite{ATLAS-750GeVExcess,CMS-750GeVExcess}
which also discuss the dark matter implications of the neutral VLFs present in those models
is studied in
Refs.~\cite{Gopalakrishna:2016tku,Bhattacharya:2016lyg,DEramo:2016aee,Mambrini:2015wyu,Ge:2016xcq,Han:2015yjk,Backovic:2015fnp}.
With more data accumulated at the LHC, it appears that the earlier diphoton excess at 750~GeV was a statistical fluctuation
and is no longer significant at both ATLAS and CMS~\cite{ATLAS:2016eeo,Khachatryan:2016yec}. 

In this work, we study the prospects of a singlet VLF to be dark matter for various dark matter masses, taking a benchmark value of the $\phi$ mass of 1~TeV. 
We also present the constraint on the $h-\phi$ mixing angle ($\theta_h$) from the LHC $hh$ channel results~\cite{Aad:2015xja}, 
which is not analyzed in the references mentioned above.
Usually in the literature, only the $h$ mediated processes are included in the dark matter
direct-detection cross section calculations.
However, for small $\theta_h$ (or when there is no mixing), the $h$ mediated processes are suppressed, 
and the $\phi$ mediated process due to the $\phi gg$ and $\phi \gamma\gamma$ effective coupling induced by the VLQ
that we include here are important.
Ref.~\cite{Ge:2016xcq} does include this contribution, although in the context of scalar dark matter    
and when the dominant contribution is the Higgs-scalar mixing contribution. 

The rest of the paper is organized as follows.
In Sec.~\ref{MODELS.SEC} we present a model with a gauge-singlet vector-like fermion dark matter,
that also contains a singlet scalar and a vector-like quark.
We present a scenario that leads to a tiny singlet-Higgs mixing,
in which case the loop induced couplings we include in this work become significant. 
We present the formulas for the SM fermion (SMF) and VLF contributions to the 
scalar-gluon-gluon ($\phi gg$) and scalar-photon-photon ($\phi\gamma\gamma$) loop-level couplings.
We compute the dominant $\phi$ decay modes. 
We infer the perturbative unitarity constraints on the $\phi$ couplings to the VLFs.
In Sec.~\ref{LHCPheno.SEC} we compute expressions for $\phi$ production in gluon-gluon fusion,
discuss the direct LHC constraints on the model, including from the di-Higgs channel, and present the LHC diphoton rate.
In Sec.~\ref{DMPheno.SEC} we present the preferred regions of parameter-space of the model that give the
correct dark matter relic-density and are consistent with direct detection constraints, also showing the future prospects. 
In Sec.~\ref{DisCon.SEC} we offer our conclusions.
In App.~\ref{AArange.SEC} we
show the range of possible diphoton rates by saturating the upper bound from the perturbative unitarity constraint, 
and also present diphoton rates in terms of the $\phi gg$ and $\phi\gamma\gamma$ effective couplings.

%%%%%%%%%%%%%%%%%%%%%%%%%%%%%%%%%%%%%%%%%%%%%%%%%%%%%%%%%%%%%%%%%
\section{Vector-like fermion dark matter model}
\label{MODELS.SEC}

A VLF is composed of two different Weyl fermions as its $L$ and $R$ chiralities that belong to conjugate representations of the gauge group. 
In contrast to this, a chiral fermion contains a Weyl fermion without its conjugate representation partner.
A gauge-singlet vector-like fermion again contains two different singlet Weyl fermions in contrast to a Majorana fermion which contains one.
For a VLF, due to the presence of both chiralities, a mass term can be written in a gauge-invariant way without involving a Higgs field.
This allows us to add TeV-scale mass terms for the VLFs.
For VLFs, fermion number is a conserved quantity.

For us, the hidden-sector is any sector that is not charged under the SM guage symmetry,
and we remain agnostic to the possibility that there are new symmetries in this sector that may even be gauged.
For example, in theories with the factor group structure, ${\cal G}_{SM} \otimes {\cal G}_{BSM}$,
where ${\cal G}_{SM}$ is the SM gauge group $SU(3)\otimes SU(2) \otimes U(1)$, and ${\cal G}_{BSM}$ is any new-physics group, 
the states charged only under ${\cal G}_{BSM}$ and singlets under ${\cal G}_{SM}$ will look like a hidden-sector to us.
To include the possibility of the hidden sector scalar $\Phi$ to be in a nontrivial representation of
${\cal G}_{BSM}$, we take $\Phi$ to be complex, with the real component denoted as $\phi/\sqrt{2}$.
For example, Ref.~\cite{Gopalakrishna:2009yz} discusses a model in which ${\cal G}_{BSM}$ is a $U(1)$ gauge symmetry.

Here, we present a model with a SM gauge-singlet hidden-sector containing a vector-like fermion
dark matter candidate $\psi$ with mass $M_\psi$, 
and a CP-even scalar $\phi$ with mass $M_\phi$, that couples to the visible SM sector via loop-induced couplings due to
an SU(2)-singlet color triplet VLQ $U$ having hypercharge $2/3$ and mass $M_U$.
(The color-triplet is the fundamental representation of the gauged $SU(3)_c$ of the SM.)
This representation of the VLQ is just one choice out of many possible, and we take this for definiteness and to explore the phenomenology.

For a {\em thermal} dark matter candidate, the hidden-sector dark matter must couple to the visible SM sector by some operators.
Some possibilities already considered in the literature include communication via:
(a) an abelian gauge boson in the hidden sector mixing with the SM hypercharge gauge boson (see for example Ref~\cite{Gopalakrishna:2008dv} and references therein); 
(b) mixing between the $\phi$ and the SM Higgs ($h$), commonly called the 'Higgs-portal' scenario (see for example Ref.~\cite{Gopalakrishna:2009yz} and references therein). 
Here, we add another possibility (c) in which the communication between the hidden sector and the visible sector is mediated by a hidden sector scalar $\phi$
with loop induced couplings to the SM. 
The $\phi$ directly couples to the dark matter at tree-level, and at loop-level to the SM, in particular to two gluons and two hypercharge gauge bosons, 
induced by a vector-like quark (VLQ) $U$.
The loop-level coupling of the $\phi$ to two hypercharge gauge bosons imply $\phi \gamma\gamma$, $\phi Z\gamma$ and $\phi ZZ$ couplings.
The $U$ is the only new state that is charged under both ${\cal G}_{SM}$ and ${\cal G}_{BSM}$ and serves to connect the two sectors
when the scalar-Higgs mixing is small,
leading to an acceptable dark matter phenomenology. 
In this work, we do not explore option (a), and present a model in which (b) and (c) are both present. 
We show that in this model, the recent large hadron collider (LHC) di-Higgs channel constraints limits the Higgs-singlet mixing in possibility (b) to be small,
and therefore including the loop-induced couplings of the $\phi$ to the SM, as in (c), will be important.
Interestingly, in this model, the visible and hidden sectors do not decouple in the limit of the Higgs-portal mixing going to zero 
since the loop-level
couplings induced by the VLQ remain as couplings between the two sectors.
We explore this limit also.

The Lagrangian of the model is~\footnote{This model and the couplings to the VLF parallels the SVU model of Ref.~\cite{Gopalakrishna:2015wwa},
and in the notation of that paper, this model may be termed as the $SVU\psi$ model.}
\bea
\Lagr_{SM} &\supset& \mu_h^2 H^\dagger H - \lambda_H (H^\dagger H)^2 \nonumber \\
          & & - \left(y_u\, \bar{q}_L \cdot H^* u_R + y_d\, \bar{q}_L H d_R
                   + y_\nu\, \bar{\ell}_L \cdot H^* \nu_R + y_e\, \bar{\ell}_L H e_R + {\rm h.c.} \right) \nonumber \ , \\ 
\Lagr_{BSM} &\supset&  \mu_\phi^2 \Phi^\dagger \Phi 
-\lambda_\phi (\Phi^\dagger \Phi)^2 - \kappa \Phi^\dagger \Phi H^\dagger H 
-\left(\mu\, \Phi H^\dagger H + {\rm h.c.} \right) \nonumber \\
& & - M_\psi \bar\psi \psi - M_U \bar{U} U
- \left(  y_\psi \Phi \bar\psi \psi + y_U \Phi \bar U U + {\rm h.c.}\right) 
\ ,
\label{SVUpsiLagr}
\eea
where we show only the relevant terms in $\Lagr_{SM}$ and do not repeat all the SM terms,
and the ``$\cdot$'' represents the anti-symmetric product in $SU(2)$ space.
We have also not shown possible $\Phi$ and $\Phi^3$ operators since they do not affect the phenomenology
being studied here. 
The $q_L, u_R, d_R$, $\ell, e_R, \nu_R$ are the 3-generation SMFs, and we suppress the generation index on these fields.
Here we have included right-handed neutrinos ($\nu_R$) also for completeness; whether this is present in nature is still being probed in experiments. 
The VLQ $U$ is in the fundamental of $SU(3)_c$ and has EM charge $+2/3$, and thus has gauge interactions
with the gluons ($g_\mu$) and hypercharge gauge bosons ($B_\mu$) exactly as the SM up-type right-chiral quarks, and are not shown explicitly.
The $\psi$, being an SM gauge singlet, has no SM gauge interactions. 
We have demanded that $\Lagr$ respect a $Z_2$ symmetry under which only $\psi$ is odd (i.e. $\psi \to -\psi$)
and all other fields are even.
This leads to an absolutely stable $\psi$ which we identify as our dark matter candidate. 
This $Z_2$ symmetry forbids the $\bar \psi \ell_L \cdot H$ operator (where $\ell_L$ is the SM lepton doublet)
which would have otherwise been allowed and caused the $\psi$ to decay.\footnote{If the $Z_2$ symmetry is not imposed,
  this operator would be allowed and is 
  the neutrino Yukawa operator, and $\psi_R$ can then be identified as the right-handed neutrino $\nu_R$.
  This possibility is not considered here since we are motivated by having a stable dark matter candidate, 
  but is extensively studied in the literature in the context of neutrino mass models. }

To not have a cosmologically stable colored relic, the decay of the VLQ $U$ can be ensured by allowing the mixed operators
\beq
\Lagr_{\rm U-SM} \supset -\tilde{y}_U \bar{U}_R\, q_L \cdot H  - \tilde m \bar{u}_R U_L + {\rm h.c.} \ ,
\label{LU-SM.EQ}
\eeq
where $q_L$ is a left-chiral SM quark doublet and $u_R$ a right-chiral SM quark singlet.
We chose the hypercharge of $U$ to be 2/3 to be able to write these operators in Eq.~(\ref{LU-SM.EQ})
that singly couple the $U$ to the SM, allowing it to decay into SM final states, thus preventing a stable colored relic.
(The same objective can be achieved by taking a hypercharge assignment of $-1/3$ instead, which then allows us to couple the VLQ to a
down-type right-chiral SM quark.)
One can ensure that experimental constraints are not violated by taking $\tilde{y}_U \ll 1$ and $\tilde m \ll M_U$,
and allowing mixings with third generation quarks only~(for details see for example Ref.~\cite{Gopalakrishna:2006kr}). 

\begin{figure}
\centering
\includegraphics[width=0.32\textwidth]{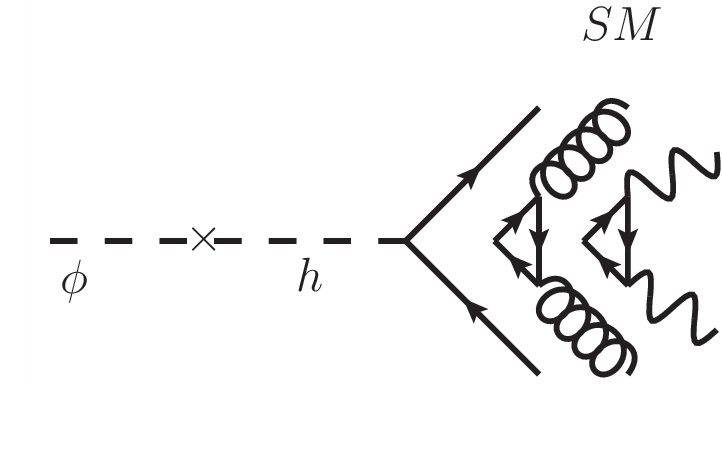}
\includegraphics[width=0.32\textwidth]{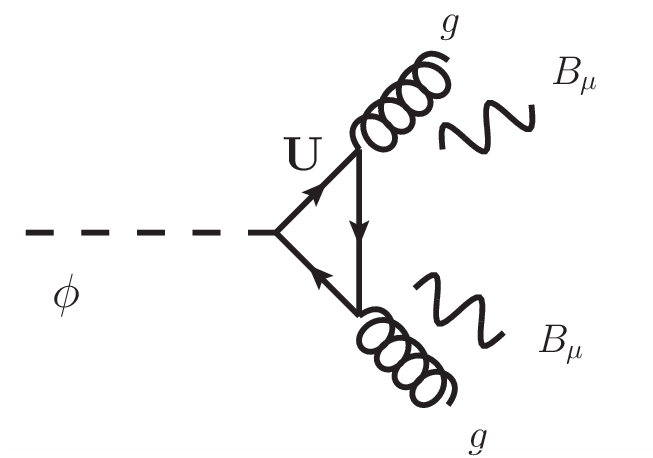}
\caption{The $\phi$ couples to the SM via the Higgs-portal mixing contribution (left),
  and via the loop-induced couplings to two gluons and two hypercharge gauge bosons due to the VLQ $U$ (right).
}
\label{phi2SM.FIG}
\end{figure}

In Fig.~\ref{phi2SM.FIG} we show schematically the two contributions to the coupling between the $\phi$
and the SM. On the left we show the Higgs-portal contribution due to $h$-$\phi$ mixing, while on the
right we show the loop-induced couplings to two gluons and to two hypercharge gauge bosons ($B_\mu$)
due to the VLQ $U$.
The latter coupling implies the $\phi\gamma\gamma$ coupling that leads to the diphoton signature explored in Sec.~\ref{LHCPheno.SEC}.

We can contemplate other hypercharge assignments for the $U$, even a hypercharge neutral assignment
with $U$ being an electroweak singlet and only charged under $SU(3)_c$.
In this case, the $U$ cannot be singly coupled to the SM since the operators in Eq.~(\ref{LU-SM.EQ}) cannot be written down,
and therefore the theory will have to be extended to allow the $U$ to decay.
We do not develop this possibility any further, other than to state that this assignment will remove the diphoton signature in Sec.~\ref{LHCPheno.SEC},
but the dark matter phenomenology of Sec.~\ref{DMPheno.SEC} will remain unchanged since that only relies on the $\phi g g$ effective coupling.

Next, we study the $\phi \leftrightarrow h$ mixing that leads to a communication between
the hidden sector and the visible SM sector, the Higgs-portal scenario.
We point out a scenario in which this mixing is suppressed. 
Following this, we work out the 1-loop $\phi gg$ and $\phi \gamma\gamma$ couplings induced by the VLQ $U$.

%%%%%%%%%%%%%%%%%%%%%%%%%%%%%%%%%%%%%
\subsection{Higgs-scalar mixing}

\begin{figure}
\centering
\includegraphics[width=0.32\textwidth]{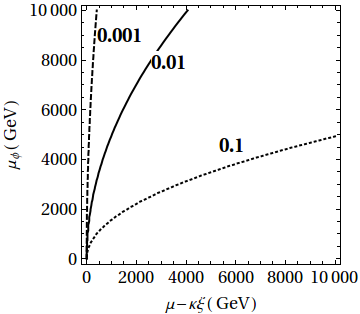}
\caption{The scalar-Higgs mixing parameter $s_h = 0.001, 0.01, 0.1$ contours as a function of the Lagrangian parameters of the model.
}
\label{shParam.FIG}
\end{figure}

If the scalar potential is such that nonzero vacuum expectation values (VEVs) are generated, namely
$\left< \Phi \right> = \xi/\sqrt{2}$ and $\left< H \right> = (0, v/\sqrt{2})^T$,
and the fluctuations around these are denoted as $\hat \phi/\sqrt{2}$ and $(0, \hat h/\sqrt{2})^T$ respectively,
the $\hat\phi$ and $\hat h$ mix due to spontaneous symmetry breaking.
The $\hat\phi \hat h \hat h$ interaction strength in Eq.~(\ref{SVUpsiLagr}) is given by $(\mu + \kappa\xi)/2$. 
Diagonalizing the $\hat\phi \leftrightarrow \hat{h}$ mixing terms,
we go from the $(\hat h, \hat\phi)$ basis to the mass basis $(h,\phi)$, 
and define the mass eigenstates to be $h = c_h \hat h - s_h \hat \phi$ and $\phi = s_h \hat h + c_h \hat \phi$
with mass eigenvalues $M_h, M_\phi$ respectively. 
The mixing angle $\sin\theta_h \equiv s_h$ is given by
\beq
\tan(2\theta_h) = \frac{2 (\mu + \kappa\xi) v}{\left(\mu_\phi^2 - \mu_h^2 \right)}   \ .
\label{t2th.EQ}
\eeq 
In Fig.~(\ref{shParam.FIG}) we show the regions of parameter space that result in a small $s_h$.
We show $s_h = 0.001, 0.01, 0.1$ contours in the $(\mu + \kappa \xi)$~--~$\mu_\phi$ plane.
In our numerical analysis below, we treat $s_h$ as an input parameter, and one can always relate it to the $\Lagr$
parameters if needed using Eq.~(\ref{t2th.EQ}).
The phenomenology due to the $\kappa$ operator is discussed in detail for example in Ref.~\cite{Gopalakrishna:2009yz}.
In the $(h, \phi)$ mass basis we have  
\beq
    {\cal L}_{\phi h h} = - \kappa_{\phi h h} \frac{M_\phi}{2\sqrt{2}} \phi h h \ ,
\label{lagrphihh.EQ}
\eeq
where we have defined the dimensionless coupling
$\kappa_{\phi h h} \equiv \tan{2\theta_h} (c_h^3 - 2 c_h s_h^2) (M_\phi^2 - M_h^2)/(\sqrt{2} v M_\phi)$.
We identify the mass eigenstate $h$ as the 125~GeV Higgs boson discovered at the LHC. 

We identify here a scenario in which $s_h \ll 1$, implying a suppressed Higgs-portal coupling.  
Consider the situation when $\mu$ is either very small or zero, and $\xi = 0$.
The former is the case when $\Phi$ has non-zero charge in ${\cal G}_{BSM}$ (see for example Ref.~\cite{Gopalakrishna:2009yz}),
and the latter when there is no spontaneous symmetry breaking in the $\Phi$ sector, 
due to taking a positive mass-squared term for $\Phi$ in the potential rather than the negative mass-squared term shown in Eq.~(\ref{SVUpsiLagr}).
In such a case, although it is broken, it is useful to consider another discrete symmetry, which we call $Z_2^\prime$, under which the $\Phi$ is odd and all other fields are even. 
The full discrete symmetry under consideration then is $Z_2 \times Z_2^\prime$, where the former $Z_2$ being exact is what is keeping the dark matter absolutely stable. 
The consequence of the $Z_2^\prime$ symmetry is that if $\mu$ is zero at some scale, then $s_h = 0$ at the tree-level at that scale, as can be seen from Eq.~(\ref{t2th.EQ}). 
The $Z_2^\prime$ however is broken explicitly by the $y_U$ and $y_\psi$ operators, and will generate the $\mu$ term
at loop-level (in fact at 3-loop). This will result in a tiny $\mu \sim 10^{-6} M_\phi$,
and $s_h$ also correspondingly small. 
This serves as an example of a scenario in which $s_h \ll 1$.
When $s_h$ is suppressed, the loop induced couplings of the $\phi$ to the SM due to the VLQ $U$ becomes important to include. 
We discuss these loop-induced couplings next.

%%%%%%%%%%%%%%%%%%%%%%
\subsection{The $\kappa_{\phi \gamma \gamma}$ and $\kappa_{\phi gg}$ loop-level effective couplings}
\label{KphiVV.SEC}

When $s_h$ is small (of the order of $0.01$), the loop induced couplings
of the $\hat\phi$ to the SM induced by exchange of the VLQ $U$ will become important.
The $\hat\phi\bar\psi\psi$ tree-level coupling and these loop induced $\kappa_{\phi gg}$ and $\kappa_{\phi \gamma\gamma}$ couplings
will then couple the dark matter VLF $\psi$ to the SM. 
The $\kappa_{\phi gg}$ and $\kappa_{\phi \gamma\gamma}$ effective couplings induced by the VLF are detailed in 
Ref.~\cite{Gopalakrishna:2015wwa}. Here we summarize these contributions for easy reference.

\begin{figure}
\centering
\includegraphics[width=0.32\textwidth]{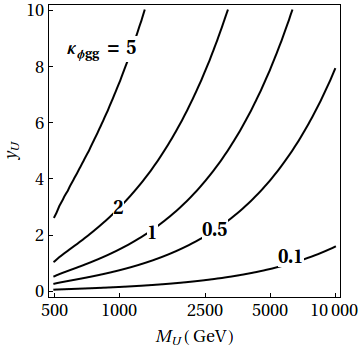}
\includegraphics[width=0.32\textwidth]{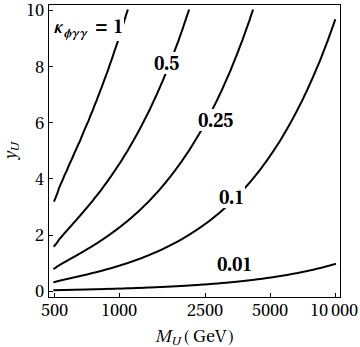}
\caption{
The $\kappa_{\phi gg}$ (left) and $\kappa_{\phi \gamma\gamma}$ (right) for $M_\phi = 1000~$GeV with $M=1000~$GeV. 
}
\label{kphiggAA.FIG}
\end{figure}

The effective Lagrangian defining the effective couplings $\kappa_{\phi gg}$ and $\kappa_{\phi \gamma\gamma}$ 
can be written for the CP-even $\phi$ following the general definitions in Ref.~\cite{Gopalakrishna:2015wwa} as
\beq
\Lagr_{\rm eff} = -\frac{\kappa_{\phi \gamma\gamma}}{64\pi^2 M} \phi F_{\mu\nu} F^{\mu\nu}
                - \frac{\kappa_{\phi gg}}{64\pi^2 M} \phi G_{\mu\nu} G^{\mu\nu}  \ ,
\label{phiggAAeffL.EQ}
\eeq
where $F_{\mu\nu}, G_{\mu\nu}$ are the photon and gluon field-strengths respectively,  
$M$ is an arbitrary mass scale which we introduce to make the $\kappa_{\phi \gamma \gamma}$ and $\kappa_{\phi gg}$
effective couplings dimensionless, and we show the numerical results of these effective couplings for $M=1~$TeV.
This choice is motivated by the presence of new physics at around the TeV scale in our model. 
The observables do not depend on $M$ since it cancels out of expressions for all observables,
as can be verified easily.
We compute these effective couplings for the model Lagrangian defined in Eq.~(\ref{SVUpsiLagr}) at 1-loop.
Defining $r_{f} = m_f^{2}/P^2$, with $P^2$ the invariant-mass-squared of the scalar, 
$f$ running over all colored fermion species (includes SMFs and VLFs) with mass $m_f$ and Yukawa couplings $y_f$, 
and with the electric charge of the fermion ($f$) denoted by $Q_{f}$,
the $\kappa_{\phi gg}$ and $\kappa_{\phi \gamma\gamma}$ at 1-loop are
(for details see Ref.~\cite{Gopalakrishna:2015wwa})
\bea
\kappa_{\phi \gamma \gamma} = 2 e^2 \sum_{f} N_{c}^f Q_{f}^{2} \, \frac{y_f}{\sqrt{2}} \frac{M}{m_f}  F_{1/2}^{(1)}(r_{f}) \ , \qquad
\kappa_{\phi gg} =  g_{s}^{2} \sum_{f} \frac{y_f}{\sqrt{2}} \frac{M}{m_f}  F_{1/2}^{(1)}(r_{f})  \ ,
\label{kphiag} \\
{\rm\text with} \ \ 
F_{1/2}^{(1)}(r_{f}) = 4 r_{f} \left(\int_{0}^{1}dy \int_{0}^{1-y} dx \frac{(1 - 4 x y)}{(r_{f}-x y)}\right) \ , \nonumber
\eea
The expressions for $F^{(1)}_{1/2}$ in Eq.~(\ref{kphiag}) reduce to the closed form expressions given in
Ref.~\cite{Gunion:1989we}.
The color-factor in $\kappa^{ab}_{\phi g g}$ is $C_{ab} = (1/2)\delta_{ab}$, where $a,b=\{1,...,8\}$ are the adjoint color indices. 
Computing a decay rate or cross-section by summing over $a,b$ gives
$\sum_{a,b} |C_{ab}|^2 = 8 (1/2)^2 = 2$ resulting in a color-factor of 2. 
In the numerical results below, we include this color factor in the $\kappa_{\phi g g}$ and suppress the color indices. 
Analogous expressions hold for the $h\gamma\gamma$ and $hgg$ effective couplings, and 
in our numerical analysis we include the contribution of the VLQ $U$ in addition to the usual SMF contributions.
In Fig.~\ref{kphiggAA.FIG} we show the numerical values of the 1-loop effective couplings
$\kappa_{\phi gg}$ and $\kappa_{\phi \gamma\gamma}$ generated by the VLQ $U$
for $M=1000~$GeV, $M_\phi = 1000~$GeV, $P^2 = M_\phi^2$, in the $y_U$--$M_U$ plane. 
%

%%%%%%%%%%%%%%%%%%%%%%%%%%%%%%%%%%5
\subsection{$\phi$ decay}
\label{phiDec.SEC}

In our analysis, we include the decay modes $\phi \to \psi \bar\psi, hh, gg, \gamma\gamma, t\bar t, \tau\bar\tau$,
where $\psi$ is the vector-like dark matter, while the rest are SM final states. 
The other SM decay final states are not important for our analysis.  
We write the $\phi$ total width $\Gamma_\phi$ in terms of $\kappa_\Gamma$, which we define as   
\beq
\Gamma_\phi \equiv \frac{\kappa_{\Gamma}^2}{16 \pi} M_\phi \ .
\label{GmphiKap}
\eeq
The contribution of each decay mode to $\kappa_\Gamma^2$ includes the couplings and phase-space factors relevant to that decay.
Expression for the $\Gamma(\phi \to XX)$ can be found for example in Refs.~\cite{Gopalakrishna:2015wwa,Gunion:1989we}.
For instance, for the decay $\phi \to \psi\psi$, via a Yukawa coupling $y_\psi/\sqrt{2}$,
we have a contribution ${\kappa_{\Gamma}^2}_{(\psi\psi)} = y_\psi^2 (1-4 M_\psi^2/M_\phi^2)^{3/2}$.
For the decay $\phi\to QQ$ into a quark-pair, the same formula holds but is multiplied by the color factor $N_c$. 
The $\kappa_{\phi h h}$ coupling identified in Eq.~(\ref{lagrphihh.EQ}) leads to the $\phi \to h h$ decay,  
which contributes to $\kappa_\Gamma^2$ an amount ${\kappa_{\Gamma}^2}_{(hh)} = (\kappa_{\phi h h}^2/4)(1-4 m_h^2/M_\phi^2)^{1/2}$.
For the loop-level decays $\phi\to gg$ and $\phi\to \gamma\gamma$, as detailed in Ref.~\cite{Gopalakrishna:2015wwa}, we have
${\kappa_{\Gamma}^2}_{(gg)} = 2 \kappa^2_{\phi g g}/(16\pi^2)^2 (M_\phi^2/M^2) $ and
${\kappa_{\Gamma}^2}_{(\gamma\gamma)} =  \kappa^2_{\phi \gamma\gamma}/(32\pi^2)^2 (M_\phi^2/M^2) $. 
%

%%%%%%%%%%%%%%%%%%%%%%%%%%%%%%%%%%%%
\subsection{Perturbative unitarity constraint}
\label{utrtyCon.SEC}

If the $\phi \bar{f} f$ Yukawa coupling $y_f$ for any fermion $f$ becomes very large, certain processes will violate perturbative unitarity.
Thus, demanding perturbative unitarity
will imply an upper bound on $y_f$. 
We assume that $\lambda_\phi$, $\kappa$ and $\mu/\mu_\phi$ of Eq.~(\ref{SVUpsiLagr}) are all small enough that there is no constraint on these.
Here, we take $f=\{\psi,U\}$ and obtain upper-bounds on $y_\psi$ and $y_U$,
the $\phi \bar\psi \psi$ and $\phi \bar{U} U$ Yukawa couplings defined in Eq.~(\ref{SVUpsiLagr}), from perturbative unitarity of the
$f \bar{f} \to f \bar{f}$ process at tree-level for $s \gg M_\phi^2, m_f^2$, where $s$ is one of the Mandelstam variables as usual.

The $l^{\rm th}$ partial wave $a_l$ of the elastic scattering amplitude is bounded by perturbative unitarity to be
$|a_l| \leq 1$~\cite{Agashe:2014kda,Peskin:1995ev}. 
For the process $f \bar{f} \to f \bar{f}$, 
the helicity amplitude in the limit of $s\gg M_\phi^2, m_f^2$ is given by ${\cal M}(++\to ++) \approx y_f^2/2$~\cite{Chanowitz:1978uj,Chanowitz:1978mv},
where the ``$+$'' denote the helicities of the fermions. 
The $0^{\rm th}$ partial wave amplitude is then readily written down as $a_0 \approx y_f^2/(32\pi)$. 
There is no $t$-channel contribution to this helicity configuration,
and other helicity configurations that are non-zero have similar sized amplitudes~\cite{Chanowitz:1978mv} and therefore should result in similar bounds.

Compared to considering the $f \bar{f} \to f \bar{f}$ channel for a single $f$,
a stronger bound could result from scattering channels with different initial and final state fermions,
i.e. from the ``coupled channels'' $f_1 \bar{f}_1 \to f_2 \bar{f}_2$ with $f_1,f_2 = \{ \psi, U\}$.
To find this, we consider in the basis $(\psi\bar\psi, U^\alpha \bar U^\alpha)$ (no sum on $\alpha$) with $\alpha=\{r,g,b\}$ the color index,
the $4\times 4$ coupled channel $a_0$ matrix
\beq
a_0 = \frac{1}{32\pi} \bmat y_\psi^2 & y_\psi y_U \\ y_\psi y_U & y_U^2 \mathbb{1}_{3\times 3}  \emat \ .
\eeq
The largest eigenvalue of this coupled channel matrix is $a_0^{\rm max} = (y_\psi^2 + 3 y_U^2)/32\pi$.
Applying the perturbative unitarity bound $|a_0^{\rm max}| \leq 1$ on the coupled channel corresponding to this maximum eigenvalue thus implies
\beq
(y_\psi^2 + 3 y_U^2) \leq 32\pi \ .
\label{yfUB.EQ}
\eeq
We ensure that this bound is satisfied in the numerical analysis of the following sections.

%%%%%%%%%%%%%%%%%%%%%%%%%%%%%%%%%%%%%%%%%%%%%%%%%%%%%%%%%%%%%%%%%%%%%%%%%
\section{LHC Phenomenology}
\label{LHCPheno.SEC}

The dark matter $\psi$, when produced at the LHC, will exit the detector as missing energy.
Searches are underway at the LHC to look for missing energy events above the SM background,
in which the dark matter recoils against one or more visible leptons, photons or jets.
In addition to such missing energy signatures, one can search for the other BSM particles in the
model defined in Sec.~\ref{MODELS.SEC}.
These include the singlet scalar $\phi$ and the VLQ $U$, and in this section we 
discuss the LHC signatures of these particles. 
We work in the narrow width approximation (NWA) in which we can write
$\sigma (pp \to \phi \to XX) \approx \sigma (p p \to \phi) \times BR(\phi \to XX) \equiv \sigma_\phi \times BR_{XX} $, where
$BR(\phi \to XX) \equiv \Gamma(\phi \to XX)/\Gamma_\phi$.
In this work, we only focus on the $\phi\to\gamma\gamma$ signature at the LHC, since in comparison,
the $BR(\phi \to Z\gamma, ZZ)$ in our model are typically smaller by a factor that ranges from
about 4 to 10 depending on $M_\phi$.

%%%%%%%%%%%%%%%%
%%%%%%%%%%%%%%%%%%%%%%%%%%%%%%%%%%%%%%%%
%\medskip
%\noindent \underline
\subsection{$\phi$ production at the LHC}
\label{MatchData.SEC}
We consider here $\phi$ production via the gluon-gluon fusion channel at the LHC.  
Rather than compute $\sigma (gg \to \phi)$ ourselves, we relate it to the SM-like Higgs production c.s. 
at this mass and make use of the vast literature on this by writing
\beq
\sigma(gg\to \phi) = \sigma(gg\to h^\prime) \frac{\Gamma(\phi \to g g)}{\Gamma(h^\prime \to gg)} \ ,
\label{pp2phiGmgg}
\eeq
where $h^\prime$ denotes a scalar with SM-Higgs-like couplings to other SM states with the mass varied.
We take from Ref.~\cite{Baglio:2010ae} the 14~TeV LHC $\sigma(gg\to h^\prime)$ 
for $M_{h^\prime} = 1000$~GeV and multiply by $0.9$ to get the $\sqrt{s}=13~$TeV values~\cite{deFlorian:2016spz}.
As can be inferred from Eq.~(\ref{pp2phiGmgg}) and detailed in Ref.~\cite{Gopalakrishna:2015wwa}, 
a quark $Q$ coupled to $\phi$ via a Yukawa coupling $y_Q/\sqrt{2}$ as in Eq.~(\ref{SVUpsiLagr}), 
gives a contribution to $\sigma(gg \to \phi)$ given by 
\beq
\sigma(gg \to \phi) = \sigma(gg\to h^\prime) \left|\sum_{Q} \frac{ y_Q}{y_t} \frac{F_{1/2}^{(1)}(r_Q)}{F_{1/2}^{(1)}(r_t)} \frac{m_t}{M_{Q}} \right|^2 \ ,
\label{gg2phiySq}
\eeq
where the sum over $Q$ includes all quarks, including the top-quark and VLQ contributions, 
$y_t = \sqrt{2} m_t/v$ is the top $htt$ Yukawa coupling (we ignore the effect of running this to the scale $\mu = M_\phi$), 
$F_{1/2}^{(1)}$ is defined in Eq.~(\ref{kphiag}) whose argument is
$r_Q \equiv m^2_Q/P^2$ with $P^2 = M^2_\phi$ for obtaining the on-shell $\phi$ resonant cross-section.
We include contributions from $Q=\{t,U\}$, and in the small scalar-Higgs mixing region ($s_h \lesssim 0.1$), the $U$ contribution dominates.

%%%%%%%%%%%%%%%%%%%%%%%%%%%%%%%%%%%%%%%%%%%%%%%%%%%
\subsection{LHC constraints}
\label{constr.SEC}
We discuss here the LHC constraints on the model from the $tt,\tau\tau$, dijet and di-Higgs channels.
In Fig.~2 of Ref.~\cite{Gopalakrishna:2015wwa}, constraints on the $\kappa_{\phi gg}$ from the
8~TeV LHC exclusion limits are shown.
For an SM-like Higgs $h^\prime$ with mass $1000$~GeV, 
we have $\kappa_{h^\prime g g} = 7$ with $\sigma(pp\to h^\prime) \approx 30~$fb at the 8~TeV LHC
due mainly to the top contribution.
From the $\kappa_{\phi gg}$ expression in Eq.~(\ref{kphiag}),
with $r_f = m_f^2/M_\phi^2$ here, since $P^2 = M_\phi^2$ for on-shell $\phi$ production, 
and with $BR_i = \kappa_{i}^2/\kappa_\Gamma^2$, we derive the bound
\beq
\left|\sum_{Q} \frac{ y_Q}{y_t} \frac{F_{1/2}^{(1)}(r_Q)}{F_{1/2}^{(1)}(r_t)} \frac{m_t}{M_{Q}} \right|^2 \frac{\kappa_{i}^2}{\kappa_\Gamma^2}  < \left(\frac{\kappa_{\phi g g\, (i)}^{\rm max}}{\kappa_{h^\prime g g}}\right)^2 \ ,
\label{kapLim.EQ}
\eeq
where the sum over $Q$ is as explained below Eq.~(\ref{gg2phiySq}), 
$\kappa_t^2 = N_c y_{\phi t t}^2 (1-4 r_t)^{3/2}$, and 
$\kappa_\tau^2 = y_{\phi \tau \tau}^2 (1-4 r_\tau)^{3/2}$.
The index $(i)$ runs over various channels $\{t\bar t, \tau\bar\tau, hh, gg, ...\}$ i.e. $(i) = \{t,\tau, h, g \}$, 
and we have $\kappa_{\phi g g\, (t)}^{\rm max} = 20$, $\kappa_{\phi g g\, (\tau)}^{\rm max} = 4$ (corresponding to $BR_i = 1$)
as derived in Ref.~\cite{Gopalakrishna:2015wwa}.
The limits on our model due to $\phi \to tt,\tau\tau$ will be very weak for small mixings $s_h \sim 0.07$. 
The LHC upper limit on the dijet channel at a mass of $1000~$GeV is about $30~$pb~\cite{Sirunyan:2016iap}, 
and for the sizes of cross-section and dijet BR in this model, this will be a loose constraint.
The 95\,\% CL limit on $\sigma(gg\to \phi) \times BR(\phi\to hh)$ at a resonance mass of $M_\phi = 1000~$GeV 
is about $10~$fb as can be read-out from the experimental exclusion plot in Fig.~6 of Ref.~\cite{Aad:2015xja} from the
ATLAS collaboration. (The $H$ of Ref.~\cite{Aad:2015xja} in our case is the $\phi$ decaying into $hh$.)
This translates into $\kappa_{\phi g g\, (h)}^{\rm max} = 4$ in Eq.~(\ref{kapLim.EQ}).
For the parameter ranges in our study, we find the di-Higgs limit is stringent and limits $s_h \lesssim 0.07$ for
$\kappa_\Gamma^2 \approx 0.1$ ({\it cf} Fig.~\ref{omega.svupsi} for the limits for a range of $s_h$).

Generically, in new physics models including the one under consideration here, there are shifts in the $h$ couplings to SM states, 
which are constrained by the LHC data (see for example Ref.~\cite{Ellis:2014dza}).
Once the above constraint $s_h \lesssim 0.07$ is enforced,
the constraints from the Higgs coupling measurements are satisfied. 

The precise direct limit on the mass of the VLQ $U$ depends on the BRs. 
The lower limit on the $U$ mass is presently in the $920-1000$~GeV range~\cite{ATLAS:2013ima,CMS:2013tda,Khachatryan:2015oba,ATLAS:tp-13TeV-3.2ifb,Aad:2015kqa},
For a long-lived VLQ with life-times in the range $10^{-7} - 10^{5}$~s, the bound is looser with $M_U \gtrsim 525~$GeV
being allowed~\cite{Khachatryan:2015jha,Aad:2013gva}.~\footnote{
  It may be possible to weaken the VLQ mass bound somewhat by allowing the decay $U \to t \phi^\prime$
  where $\phi^\prime$ is an SU(2) singlet and will lead to missing energy at the LHC. 
  This for example can be achieved by introducing the operators $U \phi' t^c$
  where $U$ is the VLQ and $t^c$ is the SM right-chiral top quark. 
  Due to the new decay mode, the usual assumption that the BRs into the SM final states ($bW,tZ,th$)
  sum to one fails, and the limits have to be reanalyzed.
  The BRs into the SM final states are decreased and since the new mode has substantially larger SM irreducible SM
  $t\bar t + \etmiss$ backgrounds, the VLQ lower limits should be weaker.
A detailed investigation of the implications of this proposal is beyond the scope of this work.
For instance, the model discussed in Ref.~\cite{Das:2015enc} has this possibility. 
}

%%%%%%%%%%%%%%%%%%%%%%%
\subsection{LHC diphoton rate}

From Eqs.~(\ref{GmphiKap})~and~(\ref{pp2phiGmgg}), the LHC 
$\sigma_\phi \times BR(\phi \to \gamma \gamma)$ in terms of the effective couplings can be written as 
\beq
\label{sigmaphiBrGm}
\sigma_\phi \times BR_{\gamma\gamma} = \left[ \sigma(gg\to h^\prime) \frac{ \kappa_{\phi gg}^2 }{ \kappa_{h^\prime gg}^2 } \right] \left[ \frac{1}{4} \left(\frac{\kappa_{\phi \gamma\gamma}}{16\pi^2 M}\right)^2 \frac{M_\phi^2}{\kappa_\Gamma^2} \right] \ ,
\eeq
where, as already explained below Eq.~(\ref{phiggAAeffL.EQ}), $M$ is a reference mass-scale which we take to be $1~$TeV.
The $\sigma_\phi \times BR_{\gamma \gamma}$ can be obtained from Eq.~(\ref{sigmaphiBrGm}),
and the expressions for $\kappa_{\phi gg}$, $\kappa_{\phi \gamma\gamma}$ are given in
Eq.~(\ref{kphiag}) with SMF and VLQ contributions included.
For $\kappa_{h^\prime gg}$ in Eq.~(\ref{kphiag}) only the SMF is included.
As a representative benchmark point, we present results in this section for $M_\phi = 1000~$GeV. 

\begin{figure}
\centering
\includegraphics[width=0.45\textwidth]{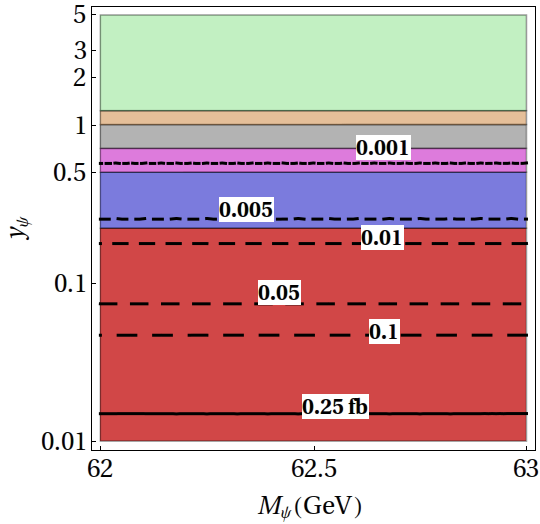}
\includegraphics[width=0.45\textwidth]{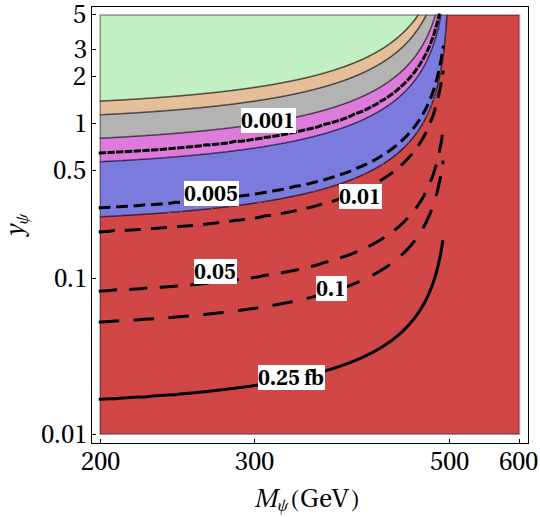}\\
\includegraphics[width=0.45\textwidth]{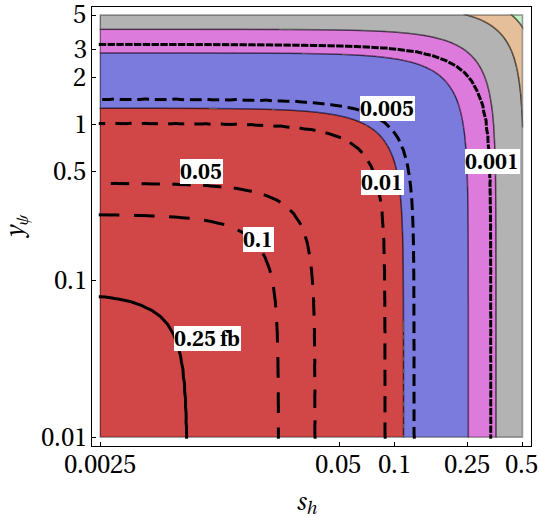}
\includegraphics[width=0.45\textwidth]{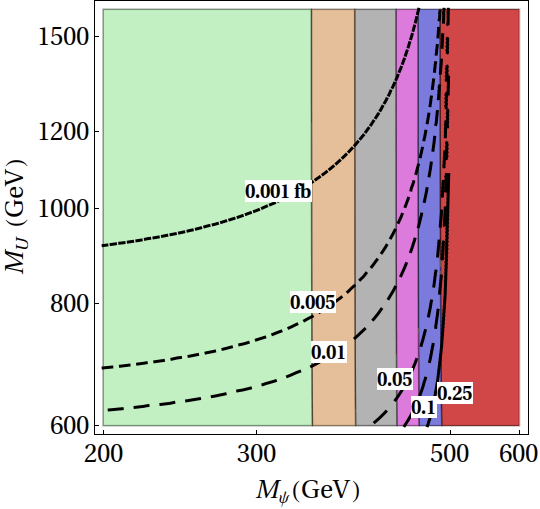}
\caption{The contours of $\sigma_\phi \times BR_{\gamma \gamma} $ (in fb), 
and regions (darker to lighter shades) of 
$\kappa_\Gamma^2 < 0.1$ (red),
$0.1<\kappa_\Gamma^2 < 0.5$ (blue),
$0.5<\kappa_\Gamma^2 < 1$ (pink),
$1<\kappa_\Gamma^2 < 2$ (gray),
$2<\kappa_\Gamma^2 < 3$ (orange),
$\kappa_\Gamma^2 > 3$ (light green);  
parameters not varied along the axes are fixed at $s_h =0.01$, 
$M_\psi = 475$~GeV, $M_U = 1200$~GeV, $M_\phi=1000$~GeV
$y_\psi = 2$, $y_U=2.5$.
These parameter choices are motivated by obtaining the observed dark matter relic density and direct-detection constraints. 
}
\label{sigmabr.svupsi}
\end{figure}

In Fig.~\ref{sigmabr.svupsi} we show contours of
$\sigma_\phi \times BR_{\gamma \gamma} $ (in fb), 
and various $\kappa_\Gamma^2$ as colored regions (darker to lighter shades correspond to smaller to larger $\kappa^2_\Gamma$), with the 
parameters not varied along the axes fixed at $s_h =0.01$, 
$M_\psi = 475$~GeV, $M_U = 1200$~GeV, $M_\phi=1000$~GeV, $y_\psi = 2$, $y_U=2.5$.
These parameter choices are motivated by obtaining the observed dark matter relic density and direct-detection constraints
({\it cf} Sec.~\ref{DMPheno.SEC}). 
For these central values of the parameters we find that $\sigma_\phi \simeq 0.25$~pb, and 
the partial widths $\Gamma_{\{hh,tt,gg\}}$ are $0.015,0.004,0.065$~GeV respectively.
The current sensitivity of the LHC searches is about $1$~fb.
For the parameters in the figure, the diphoton rate range is $0.001$-$0.25$~fb, which the LHC will probe in the future.
The entire parameter region shown in the plot satisfies the unitarity constraint in Eq.~(\ref{yfUB.EQ}).
For very small $y_\psi$ or $M_\psi > M_\phi/2$, $\Gamma(\phi \to \psi \psi) \simeq 0$ and $\Gamma_\phi$ is dominated by $\Gamma_{\{hh,tt,gg\}}$;
in this limit $BR_{\gamma \gamma} \simeq 2.4 \times 10^{-3}$ and $\sigma_\phi \times BR_{\gamma\gamma} \simeq 0.4$~fb for the set of parameters chosen with $s_h = 0.01$.
For $M_\psi < M_\phi/2$ and $y_\psi$ large, $\Gamma_\phi$ is large, being dominated by $\phi \to \psi \psi$ decay, resulting in a very small $\sigma_\phi \times BR_{\gamma \gamma}$.
In the region where $M_\psi$ is within about 5~MeV of 
$M_\phi/2$ and if $\Gamma_\psi < 0.1~$MeV, a large threshold enhancement is possible~\cite{Bharucha:2016jyr}, 
which we do not include in our analysis.

In App.~\ref{AArange.SEC}, we present model-independently the range of diphoton rates as a function of the effective couplings,
valid more generally than for the specific model considered here.
We overlay on the plots there the diphoton rate for the model considered here.
We also present the range of diphoton rate for the  model considered here by varying $y_U$ and $y_\psi$
from very small values all the way up to saturating the unitarity constraint of Eq.~(\ref{yfUB.EQ}).
While helping in probing the model considered here, these results also help more generally
in probing other such models through the diphoton channel. 

The 8~TeV $hh$ channel constraints discussed in Sec.~\ref{constr.SEC} constrains $\kappa_{\phi hh} \ll 1$. 
For example, this constraint leads to the bound $s_h \lesssim 0.17$ for $y_U = 2.5$ and $\kappa_\Gamma^2 = 0.25$.
For $y_\psi \gtrsim 0.1 $, the $BR(\phi \to \psi\psi)$ is dominant and $y_\psi$ largely controls $\kappa_\Gamma^2$. 
For $\kappa_\Gamma^2 = 1$, the $\sigma_\phi \times BR_{\gamma\gamma}$ can reach only about $0.03$~fb for $s_h = 0.01$.
For very small $y_\psi \lesssim 0.1$, the total width (i.e. $\kappa_\Gamma^2$) is small and dominated by top and $U$ loops and the tree-level $\phi \to hh,tt$ decays.
For $y_\psi \to 0$, $s_h \to 0$, both $\sigma_\phi \times BR_{\gamma \gamma}$ and $\kappa^2_\Gamma$ comes from $U$ loops and scales as $y_U^4$ and $y_U^2$ respectively;
$\sigma_\phi \times BR_{\gamma \gamma}$ increases with $\kappa^2_\Gamma$ up to around $\kappa^2_\Gamma \simeq 0.03$.

%%%%%%%%%%%%%%%%%%%%%%%%%%%%%%%%%%%%%%%%%%%%%%%%%%%%%%%%%%%%%%%%%%%%%%%%%%%%%5
\section{Dark Matter Phenomenology}
\label{DMPheno.SEC}

In this section we identify the region of parameter space of the model of Sec.~\ref{MODELS.SEC}
where the VLF $\psi$ is a viable dark matter candidate.
We also discuss constraints from dark matter direct detection experiments and prospects for the future.
Another way to probe this scenario is through indirect detection via cosmic ray observables,
which we do not take up in this study and leave for future work. 

%%%%%%%%%%%%%%%%%%%%%%%%%%%%%%%%%%%%%%%%%%%%%%%%%%%
\subsection{Dark matter relic density}
\label{dmRelDen.SEC}

%%%%%%%%%%%%%%%%%%%
The dark matter relic density is set by the self-annihilation processes $\psi \psi \to {\rm SM}$ mediated by
s-channel $h,\phi$ exchange. 
The relic density can be computed as detailed, for example, in App.~A of Ref.~\cite{Gopalakrishna:2006kr}.
We have for our case~\cite{Gopalakrishna:2009yz,Gopalakrishna:2006kr} the self-annihilation thermally averaged cross-section given by 
\beq
\left< \sigma v\right> = \frac{6}{x_f} \frac{1}{8\pi s} \sum_i  |\mathcal{B}_i|^2  \hat\Pi_{PS}^i \ ,
\label{sigv.EQ}
\eeq
where $x_f \equiv M_\psi/T_f \approx 25$ with $T_f$ the freeze-out temperature, the sum is over all self-annihilation processes $\psi\psi \to f_i f_i$ for final states $f_i$ kinematically allowed,
the $|\mathcal{B}_i|^2$ is the coefficient of $v_{rel}^2$ in the amplitude squared for each process,
$v_{rel}$ being the relative velocity of the two initial state $\psi$;
the $\hat\Pi_{PS}^i \equiv (1-4m_i^2/s)^{1/2}$ is a phase-space factor with $m_i$ the mass of the final-state particle, and $s$ is the Mandelstam variable, which
for a cold-dark matter candidate during freeze-out is $s\approx 4 M_\psi^2$.
In our analysis we include the two-body final states $b\bar b, WW, ZZ, hh, t\bar t, gg$,
whichever are kinematically allowed for that given $M_\psi$.
The loop-level $\gamma\gamma, Z\gamma$ final states are insignificant compared to $gg$, and therefore we do not include them. 
The $|\mathcal{B}_i|^2$ for each of these final states are extracted from Ref.~\cite{Gopalakrishna:2009yz}
to which we add $|\mathcal{B}_{gg}|^2$ here.
These are given by
\bea
&|\mathcal{B}_{f\bar f}|^2 =  N_c^f y_f^2 y_\psi^2 s_h^2 c_h^2 \left(1-\frac{4m_i^2}{s}\right) M_\psi^4 \hat{S}_{BW}^{h\phi} \ ; \quad 
\hat{S}_{BW}^{h\phi} = \frac{(M_\phi^2 - M_h^2)^2}{\left[ (s-M_h^2)^2 + M_h^2 \Gamma_h^2 \right] \left[ (s-M_\phi^2)^2 + M_\phi^2 \Gamma_\phi^2  \right]  } \ , \nonumber \\
&|\mathcal{B}_{WW}|^2 = \frac{y_\psi^2 g^4 v^2 s_h^2 c_h^2 M_\psi^2}{4} \left[ \frac{1}{2} + \frac{(s/2 - M_W^2)^2}{4 M_W^4} \right] \hat{S}_{BW}^{h\phi} \ , \nonumber \\
&|\mathcal{B}_{hh}|^2 = \frac{ M_\psi^2 y_\psi^2}{64} \left\{ \frac{ s_h^2 c_h^6 \kappa_{3h}^2 v^2}{\left[ (s-M_h^2)^2 + M_h^2 \Gamma_h^2 \right]} + \frac{c_h^8 \kappa_{\phi hh}^2 M_\phi^2}{\left[ (s-M_\phi^2)^2 + M_\phi^2 \Gamma_\phi^2 \right]}
  - \frac{2  s_h c_h^7 \kappa_{3h} v \kappa_{\phi hh} M_\phi}{\left[ (s-M_h^2) (s-M_\phi^2) + M_h M_\phi \Gamma_h \Gamma_\phi \right] } \right\} \nonumber \ , \nonumber \\
&|\mathcal{B}_{gg}|^2 =  \frac{16 y_\psi^2 M_\psi^6}{(16 \pi^2 M)^2}  \left\{\frac{c_h^2 \kappa_{\phi gg}^2}{(s- M_\phi^2)^2 + M_\phi^2 \Gamma_\phi^2}    + \frac{s_h^2 \kappa_{h gg}^2}{(s- M_h^2)^2 + M_h^2 \Gamma_h^2} -\frac{2 c_h s_h \kappa_{\phi gg} \kappa_{h gg} }{\left[ (s-M_h^2) (s-M_\phi^2) + M_h M_\phi \Gamma_h \Gamma_\phi \right]}\right\}  \ ,
  \label{sigvBi.EQ}
\eea
where $s\approx 4 M_\psi^2$, $\hat{S}_{BW}^{h\phi}$ is a Breit-Wigner resonance factor including the s-channel $\{h,\phi\}$ contributions, $f\bar f = \{ b\bar b, t\bar t\}$,
the ${\cal M}_{ZZ}$ is identical to ${\cal M}_{WW}$ except for an additional factor of $1/(2c_W^2)$ and $M_W \to M_Z$,
and in $|{\cal M}_{hh}|$ we do not include the $t$-channel (and $u$-channel) contributions as it is suppressed by an extra factor of $s_h$ and can be ignored for $s_h \ll 1$. 
$M$ is a mass scale which we set to $1$~TeV for numerical evaluations as explained below Eq.~(\ref{phiggAAeffL.EQ}).
We evaluate $\kappa_{\phi gg}$ and $\kappa_{h gg}$ using Eq.~(\ref{kphiag}) taking $r_f = m_f^2/(4 M_\psi^2)$,
since $P^2 = s \approx 4 M_\psi^2$ here.
The mixing angle $\theta_h$ enters in $\kappa_{\phi gg}$ and $\kappa_{h gg}$ through $\phi UU, \phi tt$ and $htt$ couplings.
Although $\tau \tau$ and $\gamma \gamma$ final states are also possible,
we neglect them in our analysis since these contributions are small owing to
a small $y_\tau$ for the former and a small EM coupling for the latter,
compared to the larger QCD coupling and the presence of a color factor in the $gg$ case.
For small $s_h \lesssim 0.07$, the $gg$ contribution becomes comparable or even larger than the tree-level contributions.

%%%%%%%%%%%%%%%%%%%%%%%%%%%%%%%%%%%%%%%%%%%%%%%%%%%
\subsection{Dark matter direct detection}
\label{dmDirDet.SEC}

The dark-matter direct-detection elastic scattering cross-section on a nucleon is
mediated by $h,\phi$ exchange.
If $s_h \lesssim 0.07$, the $\phi$ contribution is also important even though it is much heavier than $h$. 
The $h$ exchange contribution is given for example in Ref.~\cite{Gopalakrishna:2009yz}, which 
we generalize here to include $\phi$ contribution also since we consider $s_h \lesssim 0.07$. 
The scalar-nucleon-nucleon coupling is generated due to the scalar coupling to the quark content of the nucleon, 
and also due the scalar coupling to the gluon content of the nucleon via the $ggh,gg\phi$ effective couplings.  
We define an effective Lagrangian for the scalar-nucleon-nucleon interaction as
\bea
\Lagr &\supset& \frac{\lambda_{hNN}}{\sqrt{2}} \hat{h} \bar{N} N + \frac{\lambda_{\phi NN}}{\sqrt{2}} \hat{\phi} \bar{N} N \ , \nonumber \\
         &=& \frac{(c_h \lambda_{hNN} - s_h \lambda_{\phi NN})}{\sqrt{2}} h \bar{N} N + \frac{(c_h \lambda_{\phi NN} + s_h \lambda_{hNN} )}{\sqrt{2}} \phi \bar{N} N \ , 
\eea
where $N$ denotes the nucleon, and in the second line we write in the mass basis.
We derive $\lambda_{h N N}$ and $\lambda_{\phi N N}$ using the formalism of Ref.~\cite{Shifman:1978zn}
updated in Ref.~\cite{Ellis:2000ds} (for a review, see App.~C of Ref.~\cite{Bertone:2004pz}) in which the scalar-nucleon coupling is denoted as $f_{p,n}$.
Identifying $\lambda_{hNN} \equiv f_{p,n}$, we have~\cite{Ellis:2000ds}
\beq
\lambda_{hNN} = \sum_{q=u,d,s} f_{Tq}^{(p,n)} y_q \frac{m_{p,n}}{m_q} + f^{(p,n)}_{TG} \frac{2}{27} \sum_{q=c,b,t} y_q \frac{m_{p,n}}{m_q} \ ,
\label{lamhNN.EQ}
\eeq
with~\cite{Ellis:2000ds} $f_{Tu}^{(p)} = 0.02 \pm 0.004$, $f_{Td}^{(p)} = 0.026 \pm 0.005$, $f_{Ts}^{(p)} = 0.118 \pm 0.062$,
$f_{Tu}^{(n)} = 0.014 \pm 0.003$, $f_{Td}^{(n)} = 0.036 \pm 0.008$, $f_{Ts}^{(n)} = 0.118 \pm 0.062$,
and $f^{(p,n)}_{TG} = 1-\sum_{q=u,d,s} f_{Tq}^{(p,n)} \approx 0.85$.
We then find numerically $\lambda_{hNN} = 2 \times 10^{-3}$, which we take for our numerical analysis.
The $\lambda_{\phi N N}$ coupling is induced via the $\phi$ couplings to the gluon content of the nucleon, i.e. via the $\phi g g$ effective coupling induced by the VLQ $U$.
Following the same procedure as above, we derive this coupling from the second term in Eq.~(\ref{lamhNN.EQ})
as $\lambda_{\phi NN} = (2/27)\, f^{(p,n)}_{TG}\, y_U m_{(p,n)}/M_U  \approx 0.063\, y_U\, m_N/M_U$, which we use for our numerical studies. 
We content ourselves with this simple estimate of the coupling. 
Reliably computing the effective coupling is of critical importance, and 
our direct-detection rates can be scaled quite straightforwardly for a more accurately computed coupling. 
Ref.~\cite{Cheng:2012qr} examines recent developments and argues for a smaller value of the coupling $\lambda_{hNN} \approx 1.1 \times 10^{-3}$.
Other sophisticated analyses can be found for example in Refs.~\cite{DEramo:2016aee,hNNcoup.REF}. 

We can now write the spin-averaged $\psi$ elastic scattering cross section on a nucleon for $q^2 \ll m_N^2$ as 
\bea
\sigma(\psi N \to \psi N) &=& \frac{y_\psi^2}{8 \pi} 
   \left[\frac{s_h (c_h \lambda_{hNN} - s_h \lambda_{\phi NN})}{M_h^2} - \frac{c_h (c_h \lambda_{\phi NN} + s_h \lambda_{h NN})}{M_\phi^2}  \right]^2 
    \left( |{\bf p}_\psi|^2 + m_N^2 \right) \ , \nonumber \\
  &=& \frac{y_\psi^2 s_h^2 c_h^2 \lambda_{hNN}^2}{8 \pi} \frac{\left( |{\bf p}_\psi|^2 + m_N^2 \right)}{M_h^4} 
    \left[1 - \frac{\lambda_{\phi NN}}{\lambda_{h NN}} \frac{c_h}{s_h} \frac{(1+\Delta_\phi)}{(1-\Delta_h)} \frac{M_h^2}{M_\phi^2} \right]^2 \ ,
\label{dirDetCS.EQ}
\eea
where $p_\psi\approx M_\psi v_\psi$ with $v_\psi \sim 10^{-3}$~\cite{Bertone:2004pz}, $m_N \approx 1~$GeV is the nucleon mass, 
$\Delta_h = (\lambda_{\phi NN} / \lambda_{h NN}) (s_h/c_h)$, and $\Delta_\phi = (\lambda_{h NN} / \lambda_{\phi NN}) (s_h/c_h)$.
This is the generalization of the direct-detection elastic cross section Eq.~(13) of Ref.~\cite{Gopalakrishna:2009yz}
which included only the $h$ contribution, to now include the $\phi$ contribution also that becomes important
for very small $s_h$.
For instance, for $s_h = 0.01$, $M_\phi = 1000$~GeV, the extra factor in Eq.~(\ref{dirDetCS.EQ}), namely, 
$\left[...\right]^2 \approx \left[1-0.125\, (1200~{\rm GeV}/M_U) (y_U/2.5) \right]^2$, with $\Delta_{\phi, h} \ll 1$ which can be dropped. 
Including the $\phi$ contribution thus {\em decreases} the elastic cross-section by about $25$\,\% for the central values we choose.

In addition to uncertainties in the dark matter nucleon effective coupling mentioned above, 
there is uncertainty in the local dark matter halo density and its velocity distribution.
Given these uncertainties (see for example Refs.~\cite{SigmaDD.Un}), the direct-detection exclusion limits should be taken to be accurate
only up to unknown ${\cal O}(1)$ factors.

%%%%%%%%%%%%%%%%%%%%%%%%%%%%%%%%%%%%%%%%%%
\subsection{Dark matter preferred regions of parameter space}
\label{dmPrefReg.SEC}

Here we show regions of parameter space of the model of Sec.~\ref{MODELS.SEC} for 
which we obtain the observed relic-density and are consistent with the dark matter direct detection limits.
We also present the prospects in upcoming direct detection experiments.
In order to get the correct relic density of $\Omega_{dm} = 0.26 \pm 0.015$~\cite{Adam:2015rua},
we need the thermally averaged self-annihilation cross-section to be
$\left< \sigma v\right> \approx 2.3\times 10^{-9}~{\rm GeV}^{-2}$.

\begin{figure}
\centering
\includegraphics[width=0.32\textwidth]{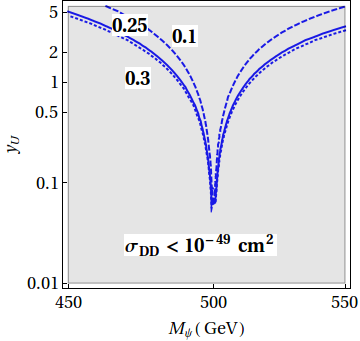}
\includegraphics[width=0.32\textwidth]{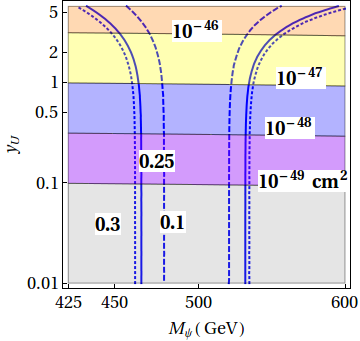}
\caption{Contours of $\Omega_{dm} = 0.1,~0.25,~0.3$, for $y_\psi=2$, $M_\phi=1000$~GeV, $M_U = 1200~$GeV,
  for $s_h = 0$ (left) and $s_h = 0.05$ (right)
  with the colored bands showing $\sigma_{DD}$ as marked. 
}
\label{OmDD-yUMpsi.FIG}
\end{figure}

\begin{figure}
\centering
\includegraphics[width=0.32\textwidth]{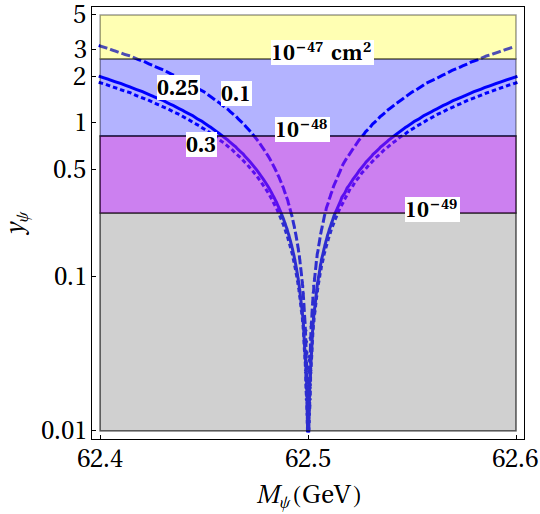}
\includegraphics[width=0.32\textwidth]{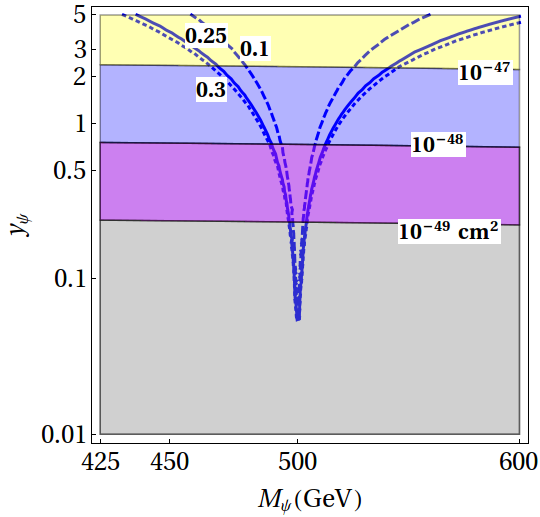}
\includegraphics[width=0.32\textwidth]{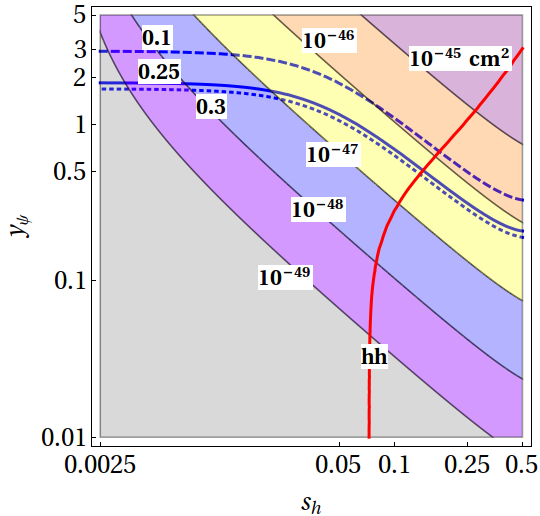}
\caption{
%  Contours of $\Omega_{dm} = 0.1,~0.25,~0.3$, for $M_\phi=1000$~GeV, $M_U = 1200~$GeV,
%  with the colored bands showing $\sigma_{DD}$ as marked,
  Same as in Fig.~\ref{OmDD-yUMpsi.FIG} but with
  the parameters not varied along the axes fixed at $y_U=2.5$, $s_h = 0.01$ and $M_\psi =475$~GeV.
  The thick red line shows the 8~TeV LHC $hh$ channel constraint.
}
\label{omega.svupsi}
\end{figure}

In Figs.~\ref{OmDD-yUMpsi.FIG}~and~\ref{omega.svupsi} we plot contours of $\Omega_{dm} = 0.1,~0.25,~0.3$
with both the loop-induced couplings due to the the $U$ and the Higgs-portal couplings present
for $M_\phi=1000$~GeV and $M_U = 1200$~GeV,
and show the regions with
$ \sigma_{DD} >  5 \times 10^{-45}~$cm$^2$,
$10^{-45}~{\rm cm}^2 < \sigma_{DD} < 5 \times 10^{-45}~$cm$^2$,
$ 10^{-46}~{\rm cm}^2 < \sigma_{DD} < 10^{-45}~$cm$^2$,
$ 10^{-47}~{\rm cm}^2 < \sigma_{DD} < 10^{-46}~$cm$^2$,
$ 10^{-48}~{\rm cm}^2 < \sigma_{DD} < 10^{-47}~$cm$^2$,
$ 10^{-49}~{\rm cm}^2 < \sigma_{DD} < 10^{-48}~$cm$^2$,
$ \sigma_{DD} < 10^{-49}~$cm$^2$
with parameters not varied along the axes fixed at $s_h=0.01$, $M_\psi = 475$~GeV, $y_U=2.5$.
The entire parameter region shown in the plots satisfies the unitarity constraint in Eq.~(\ref{yfUB.EQ}).
We see that for the choice of parameters we make,
the direct-detection cross section is
less than the current experimental limit, which is  
$\sigma_{DD} \leq (0.1 - 1) \times 10^{-45}$~cm$^2$~\cite{DirDetExptCite} for dark matter mass in the $10-1000$~GeV range.
The correct self-annihilation cross-section is obtained 
only with an enhancement of the cross-section at the $\phi,h$ pole with 
$M_\psi \sim M_{\phi,h}/2$.
Being close to the $\phi$ pole suppresses the $\phi \to \psi\psi$ decay rate due to the limited phase-space available, leading to a small $\kappa_\Gamma^2 \lesssim 0.1$
as can be seen from Fig.~\ref{sigmabr.svupsi}.

We first explore the $s_h = 0$ limit,
i.e. when the dark matter couples to the SM entirely via the $\phi gg$ and $\phi \gamma\gamma$ effective couplings induced by the VLQ at the loop-level,
with no contribution from the Higgs portal. 
This limit can be straightforwardly taken in Eqs.~(\ref{sigv.EQ})-(\ref{dirDetCS.EQ}).
The correct relic density can be achieved in this limit as shown in Fig.~\ref{OmDD-yUMpsi.FIG} (left), and we obtain $\sigma_{DD} < 10^{-49}~{\rm cm}^2$.  
The required relic-density, for example, can be obtained for $y_\psi=2$, $y_U=2.5$, $M_U=1200$~GeV, $M_\phi=1000$~GeV, $M_\psi=467$~GeV,
for which $\sigma_{DD}=2.4\times 10^{-51}~{\rm cm}^2$.
Thus the $s_h = 0$ limit provides an example scenario in which the relic density is satisfied but direct-detection is very challenging.

In Fig.~\ref{OmDD-yUMpsi.FIG} (right) we show the situation for $s_h = 0.05$, i.e. when the Higgs-portal is also turned on.
For $y_U \gtrsim 1$ the loop induced couplings due to the $U$ are significant, while for smaller $y_U$ the Higgs-portal contribution dominates. 
Thus, for $y_U \lesssim 0.5$ the relic density contour starts losing dependence on $y_U$, and 
for $y_U = 0.1$ the loop-induced couplings are completely negligible and the dark matter phenomenology is that of the Higgs-portal scenario. 

Since we are required to have $s_h \ll 1$ in which case the $gg$ contribution dominates,
the dark matter relic-density scales as $\sim (y_\psi y_U)^{-2}$ to a very good approximation as can be inferred from 
Eqs.~(\ref{sigv.EQ}) and (\ref{sigvBi.EQ}).
Similarly, the dark matter direct-detection rate also scales the same way in this limit, 
as evident from Eq.~(\ref{dirDetCS.EQ}).  
Thus, for $M_\phi = 1000$~GeV and for a given value of $M_\psi$, 
other values of $(y_\psi,y_U)$ that give the correct relic-density 
and direct-detection rates can be obtained from those in Fig.~\ref{omega.svupsi}, by scaling $y_\psi \to (2.5/y_U) y_\psi$.
Thus, for $s_h \ll 1$, 
since the couplings of the dark matter with SM states is via loop-level effective couplings, 
we find for $M_\phi = 1000~$GeV and $M_\psi = 475~$GeV, 
moderately large values $y_\psi y_U \approx 5$ are required in order for the 
dark matter self-annihilation cross-section to be of sufficient size to give the correct relic-density.
Taking smaller values of $y_\psi y_U$ will require tuning $M_\psi$ closer to $M_\phi/2$ (or to $M_h/2$).
The regions we identify are safe from present direct-detection constraints, and will be probed in upcoming experiments.

%%%%%%%%%%%%%%%%%%%%%%%%%%%%%%%%%%%%%%%%%%%%%%%%%%%%%%%%%%%%%%%%%%%%%%%%%%%
\section{Conclusions}
\label{DisCon.SEC}
In this work, 
we study a BSM model with a hidden sector containing a stable gauge-singlet vector-like fermion dark matter $\psi$, and a gauge-singlet scalar $\phi$.
The $\phi$ couples to the SM
via its mixing to the Higgs (the Higgs-portal scenario),
and via loop-level couplings to two gluons and also to two hypercharge gauge bosons
induced by an SU(2) singlet vector-like quark $U$ carrying hypercharge $2/3$. 
We point out a scenario in which the Higgs-portal mixing is suppressed,
due to which the loop-level couplings are the dominant communication mechanism between the hidden sector and the SM.
We study the LHC and dark matter phenomenology of this model. 

We highlight the LHC direct constraints relevant to the model.
We show that the LHC di-Higgs channel constrains the Higgs-singlet mixing to be very small ($\sin{\theta_h} \lesssim 0.07$), 
and therefore the loop-induced couplings are important to include.
We present the rate for LHC scalar production via gluon-gluon fusion and its decay into the diphoton channel. 
We identify viable regions of parameter-space where the observed dark matter relic density is obtained and
that are consistent with dark matter direct detection constraints. 

When the mixing is tiny, and the dark matter is coupled to the SM via loop-induced operators,
we show that moderately large $\phi$ Yukawa couplings to the vector-like fermions $y_\psi y_U \approx 5$ are required
in order to get a large enough dark matter self-annihilation cross section to obtain the correct relic density.
Furthermore, $(M_\psi, M_\phi)$ needs to be in the pole enhanced region, i.e. $M_\psi$ should be within a few tens of GeV of $M_\phi/2$ 
(or a few tenths of GeV of $M_h/2$). 
We show that these large couplings are within the bounds of perturbative unitarity,
by computing the upper bounds on these couplings from the $f\bar{f} \to f\bar{f}$ coupled channel scattering process for $f=\{\psi,U\}$.

The diphoton rate when the scalar-fermion couplings are varied is shown in Fig.~\ref{sigmabr.svupsi}.
These diphoton rates are accessible at the LHC. 
We find regions of parameter space that are compatible with dark matter direct-detection bounds, 
and the rate we find is accessible in current and upcoming direct-detection experiments.
We show these in Fig.~\ref{omega.svupsi}, with the region consistent with the direct LHC $hh$ bound.
In addition to the direct production signals of the vector-like quark at the LHC, 
another promising signal is the $\phi\to h h$ mode which already imposes very tight constraints on the parameter-space. 
For the benchmark values of the parameters we study, the regions that yield the correct dark matter relic density
have direct detection cross-sections that range between the current limits from experiments to about $10^{-51}~{\rm cm}^2$. 
The lower value, very challenging to experimentally detect,
is obtained when the Higgs-portal mechanism is shut-off with the dark matter coupled to the SM only via the loop-level couplings.

%%%%%%%%%%%%%%%%%%%%%%%%%%%%%%%%%%%%%%%%%%%%%%%%%%%%%%%%%%%%%%%%%%%%%%%%%%%
\appendix
%%%%%%%%%%%%%%%%%%%%%%%%%%%%%%
%%%%%%%%%%%%%%%%%%%%%%%%%%%%%%%%%%%%%%%%%%%%%
% Put the following in place of the \appendix command
%%%%%%%%%%%%%%%%%%%%%%%%%%%%%%%%%%%%%%%%%%%%%
%\setcounter{section}{0}
%\renewcommand\thesection{Appendix \Alph{section}}               % use this to modify the section numbering style
%\renewcommand\thesection{\Alph{section}}               % use this to modify the section numbering style
%\renewcommand\thesubsection{\Alph{section}.\arabic{subsection}}
%\renewcommand\thesubsubsection{\Alph{section}.\arabic{subsection}.\arabic{subsubsection}}

%\renewcommand{\theequation}{\Alph{section}.\arabic{equation}}    % use this to modify the equation numbers
%\renewcommand{\thetable}{\Alph{section}.\arabic{table}}          % use this to modify the table numbers
%\renewcommand{\thefigure}{\Alph{section}.\arabic{figure}}        % use this to modify the figure numbers

%%%%%%%%%%%%%%%%%%%%%
\section{Range of diphoton rate}
\label{AArange.SEC}

\begin{figure}
\centering
\includegraphics[width=0.32\textwidth]{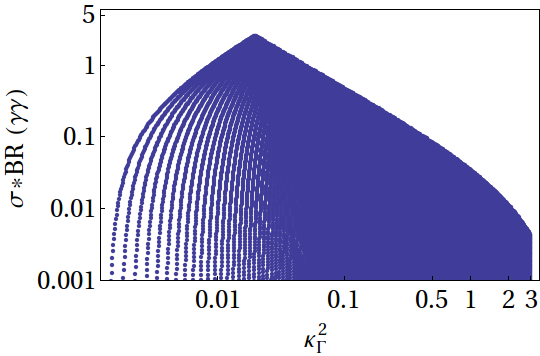}
\includegraphics[width=0.32\textwidth]{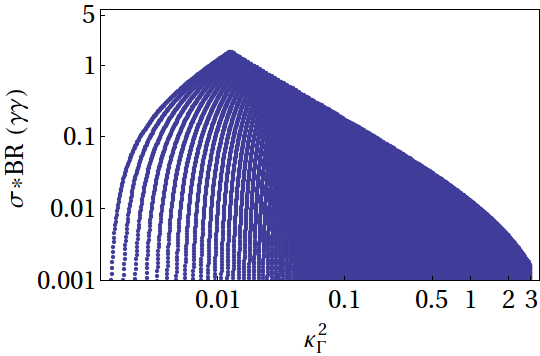}
\caption{For the model of Sec.~\ref{MODELS.SEC}, the $\sigma_\phi \times BR_{\gamma \gamma}$ (in fb) vs. $\kappa_\Gamma^2 $,
  for $M_\psi =475~$GeV, $M_U= 1200~$ (left) and $1500~$GeV (right), $M_\phi=1000$~GeV, $s_h=0.01$
  and $y_U, y_\psi$ scanned over the range  $0 < y_U < y_U^{max}$, $0 < y_\psi < y_\psi^{max}$  subject to the unitarity constraint
  in Eq.~(\ref{yfUB.EQ}).
}
\label{sigmabrScan.svupsi}
\end{figure}

In Sec.~\ref{utrtyCon.SEC} we derived an upper bound on $y_\psi$ and $y_U$ from perturbative unitarity. 
Here, we show the range of diphoton rate at the LHC by saturating this upper bound.  
In Fig.~\ref{sigmabrScan.svupsi} we show $\sigma_\phi \times BR_{\gamma \gamma}$ vs. $\kappa_\Gamma^2$ in the model of Sec.~\ref{MODELS.SEC},
for $M_\psi =475~$GeV, $M_U= 1200~$ and $1500$~GeV, $M_\phi=1000$~GeV, $s_h = 0.01$ and scanning over $y_U,y_\psi$
in the range $0 < y_U < y_U^{max}$, $0 < y_\psi < y_\psi^{max}$, subject to the unitarity constraint
in Eq.~(\ref{yfUB.EQ}).
For example, for $s_h=0.01$, we can get $\sigma_\phi \times BR_{\gamma \gamma} \simeq 2.9$~fb.

%%%%%%%%%%%%%%%%%%%%%%%%%%%%%%%%%%%%%%%%%%%%%%%%%%%%55

\begin{figure}
\centering
\includegraphics[width=0.32\textwidth]{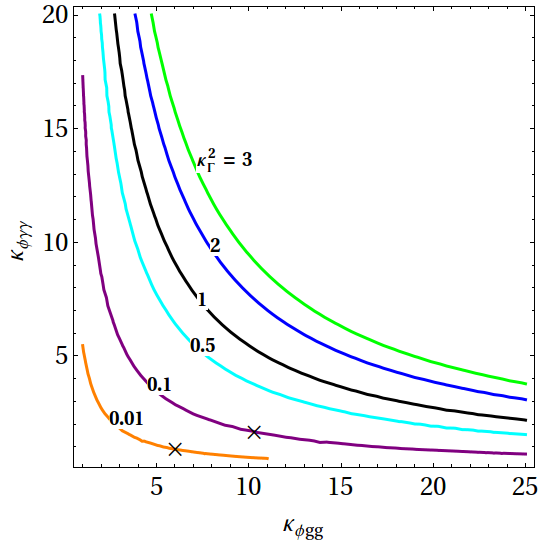}
\includegraphics[width=0.32\textwidth]{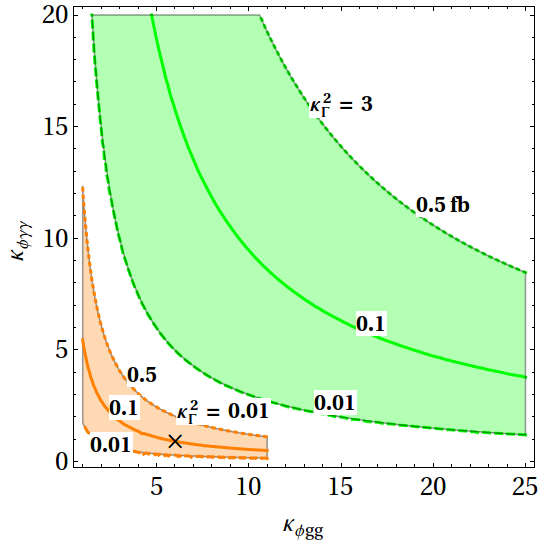}
\caption{For various $\kappa^2_\Gamma$ shown, the $\kappa_{\phi gg}$ and $\kappa_{\phi \gamma\gamma}$ 
that leads to $\sigma_\phi \times BR_{\gamma\gamma} = 0.1~$fb (left),
and the regions $0.01 \leq \sigma_\phi \times BR_{\gamma\gamma} \leq 0.5$~fb (right)
for $\kappa^2_\Gamma = 0.01, 3$~(orange and green respectively) with $M_\phi=1000$~GeV,
for the choice of the reference mass scale $M=1$~TeV.
The plots apply to any model with a $\phi$ as here,
with the ``$\times$'' showing the values for the particular model of Sec.~\ref{MODELS.SEC}
with parameter values as listed in the text.
}
\label{kggkaaAl}
\end{figure}

In Fig.~\ref{kggkaaAl} we show contours of various $\kappa^2_{\Gamma}$ in the $\kappa_{\phi gg}$--$\kappa_{\phi \gamma\gamma}$ plane
that give $\sigma_\phi \times BR_{\gamma\gamma} = 0.1$~fb for $M_\phi=1000$~GeV.
This cross-section is presently allowed with the $95\%$ CL exclusion limit being about $1$~fb~\cite{ATLAS:2016eeo,Khachatryan:2016yec}. 
We show in Fig.~\ref{kggkaaAl} (right)
a band of diphoton rate $0.01 \leq \sigma_\phi \times BR_{\gamma\gamma} \leq 0.5$~fb
for two representative total width values $\kappa^2_\Gamma = {0.01 \ {\rm and}\ 3}$, a wide range with the former being $0.1$\% of $M_\phi$, and the latter $5\%$. 
The latter width is rather large, and for $M_U > M_\phi/2$, it is obtained
for $M_\psi < M_\phi/2$ for large $y_\psi$ as we discuss below.
For such large couplings, there is a danger of tree-level unitarity being violated,
and our analysis of Sec.~\ref{utrtyCon.SEC} becomes relevant.
Fig.~\ref{kggkaaAl} shows the situation model independently in any model with a $\phi$ as here,
in which the $\kappa_{\phi gg}$ and $\kappa_{\phi \gamma\gamma}$ effective couplings can be calculated.
In the same figure, we overlay a ``$\times$'' to depict the situation for the particular model of Sec.~\ref{MODELS.SEC},
with the choice $M_U = 1200$~GeV, $M_\psi = 475$~GeV and $(y_U,y_\psi) = (2.3,0.4)$ for $\kappa^2_\Gamma=0.01$,
and $(4,1.3)$ for $\kappa^2_{\Gamma}=0.1$.

%%%%%%%%%%%%%%%%%%%%%%%%%%%%%%%%%%%%%%%%%%

\end{document}